# Relative Rate Reduction Based Control with Adjustable Congestion Level

Péter Hága, Ferenc Tóth
Communication Networks Laboratory, Eötvös University
Budapest, Hungary
E-mail: {haga, feri}@elte.hu

István Csabai, Gábor Vattay
Collegium Budapest – Institute for Advanced Study
Budapest, Hungary
Email: {csabai, vattay}@colbud.hu

*Abstract*—In Future Internet it is possible to change elements of congestion control in order to eliminate jitter and batch loss caused by the current control mechanisms based on packet loss events. We investigate the fundamental problem of adjusting sending rates to achieve optimal utilization of highly variable bandwidth of a network path using accurate packet rate information. This is done by continuously controlling the sending rate with a function of the measured packet rate at the receiver. We propose the relative loss of packet rate between the sender and the receiver (Relative Rate Reduction, RRR) as a new accurate and continuous measure of congestion of a network path, replacing the erratically fluctuating packet loss. We demonstrate that with choosing various RRR based feedback functions the optimum is reached with adjustable congestion level. The proposed method guarantees fair bandwidth sharing of competitive flows. Finally we present testbed experiments to demonstrate the performance of the algorithm.

## I. INTRODUCTION

In recent years new challenges appeared due to several developments in the networking technology. They include the larger heterogeneity of both the physical infrastructure and the applications. Network architecture spans from multi-gigabit connections to wireless and mobile connections, or even to sensor networks with limited bandwidth. Concerning applications, the typical web traffic requires transferring small amount of data quickly, while for downloading large science archives the momentary speed of data transfer is not so important. For media streaming applications other requirements appears, like smooth transfer with limited jitter.

To cope with these requirements a transport protocol has to fulfill the following goals. The physical and available bandwidths and other network resources must be used efficiently, while it is essential to react and adapt to the changing conditions quickly. The intra-protocol fairness must be guaranteed by the algorithms themselves without any explicit information or intervention from network elements. With other kinds of protocols and flows friendly behaviour is expected, a share in the network resources have to be provided. Wireless environments require robustness against single or multiple packet losses no matter whether they are lost due to fading or buffer overload.

The general traffic control in the current Internet is dominated by TCP variants [1]. Since the birth of the original concept of TCP [2] there have been significant improvements of the algorithm. Beyond the general purpose improvements [3], [4] there are modifications focusing on special environments like high bandwidth delay product connections [5]–[7] or wireless and mobile networks [8], [9]. The packet sending strategies of these transport protocols are based on some measured and inferred parameters of the network path. There are special environments where one-way information is accessible [10], [11], but usually the clocks of senders and receivers are not synchronized and time measurements should rely on information available at one of the sides. One approach is to measure round-trip information of individual packets (e.g. the data and the corresponding acknowledgment packets). The round-trip-time (RTT) and the packet loss are the most common quantities used in congestion control algorithms. There are other important quantities such as the available bandwidth, which can be used to optimize a control algorithm, but is not directly accessible. To infer such network parameters packet spacing information of consecutive packets can be used, which also accessible without the time synchronization of end hosts [12]–[20]. Rather than using the original window-based concept of TCP, some packet spacing information and rate information are also employed in some protocol variants [21]–[23].

Although the spacing of sent and received packets and the packet rate in general can be calculated easily from the timestamps of packets at each side, the superiority of protocols exploiting this information has not been demonstrated yet. We show, that using this direct rate information a new control mechanism can be developed, which can achieve optimal utilization of highly variable available bandwidth of a network path. This is done by continuously controlling the sending rate with a function of the measured packet rate at the receiver. We introduce the relative loss of packet rate (Relative Rate Reduction, RRR) between the end hosts as a new measure of congestion in a network path. Choosing various RRR based feedback functions the congestion level in the network path can be adjusted directly. This inherently achieves fair bandwidth sharing independently of RTT, provides adjustable friendliness to other protocols and quick adaptation to available bandwidth. Since the RRR based feedback is not sensitive for packet losses the investigated congestion control algorithm is robust even against high packet loss rates.

The relation of the measured rate information and the congestion level is investigated in Section II, where we also define the Relative Rate Reduction (RRR). In Section III we introduce a congestion control algorithm based on the measured RRR. Three basic feedback functions and the bandwidth utilizations are investigated in Section IV. In Section V we evaluate the performance of the proposed method in various simulated scenarios. In Section VI the performance of the UDP-based implementation of our algorithm is tested in scenarios relevant in mobile wireless environments.

## II. MEASUREMENT BASED PROTOCOL

The measured rate information is related to the congestion in the network path directly. In case of congestion the rate of a packet train at the receiver end is lower in average than its initial value at the sender side. The physical capacity of a link is shared by the flows entering simultaneously. If the total rate of flows does not exceed the physical link capacity each flow can pass through with unchanged rate. If the total rate of flows exceeds the link capacity then congestion sets up and the received share of a flow is proportional to its input rate. The macroscopic fluid equations describing the input and output rates are

$$X^{out} = \begin{cases} X^{in} & X^{in} + X_c \leq C \\ C \cdot \dfrac{X^{in}}{X^{in} + X_c} & X^{in} + X_c > C, \end{cases} \quad (1)$$

where $C$ is the physical capacity, while $X_c$ represents the rate of all the other flows.

If congestion persists *for a sufficiently long* time interval and buffers are finite, the difference between the average output and input rates of a flow is lost and the loss probability is given by

$$p = \frac{X^{in} - X^{out}}{X^{in}}. \quad (2)$$

For a single link the output rate of a flow can be written in terms of the loss probability

$$X^{out} = (1 - p) \cdot X^{in}. \quad (3)$$

This can be easily extended to multiple links via

$$X_i^{out} = (1 - p_i) \cdot X_i^{in}, \quad (4)$$

where $p_i$ is the stationary packet loss probability on the $i$-th link and

$$X_{i+1}^{in} = X_i^{out}, \quad (5)$$

for the $i$-th and $i+1$-th links. The output rate of a flow for a network path having $n$ links is

$$X_n^{out} = \prod_i^n (1 - p_i) \cdot X_1^{in}, \quad (6)$$

where $X_1^{in}$ is the input rate at the first link. Equation (3) remains valid for the multiple link case and the total loss probability can be expressed in terms of the link losses

$$p = 1 - \prod_i^n (1 - p_i). \quad (7)$$

For *shorter* congested time intervals packets are not necessarily lost with probability (2), they can queue up in the buffers instead. The difference between the average input and output rates for short intervals is the combined result of real packet loss and the increased spacing of packets caused by the sharing of the queues with the background traffic. Equation (1) remains valid in this case as well. We can in general introduce the Relative Rate Reduction (RRR) of a flow

$$\hat{p} = \frac{X^{in} - X^{out}}{X^{in}}, \quad (8)$$

which can be written in terms of the RRR-s of the links

$$\hat{p} = 1 - \prod_i^n (1 - \hat{p}_i), \quad (9)$$

where $X_i^{out} = (1 - \hat{p}_i) \cdot X_i^{in}$. The actual value of RRR for a given train of packets is a good statistical estimator of the expected long time packet loss probability at a given packet sending rate.

The RRR has several advantages over packet loss and packet loss probability: Packet loss is a discrete event happening rarely. In order to get a reasonable loss probability estimate we have to measure it for a long time. This prevents us from detecting packet loss probability changes quickly and accurately. On the other hand, RRR can be measured instantaneously and it gives direct information on the actual state of the congestion. Its measured value is a statistical indicator of the long term packet loss probability if conditions remains unchanged. Packet loss based congestion mechanisms set back the packet sending rate immediately after the observation of packet loss. Usually the rate is reduced below the actual available bandwidth of the path. In absence of packet loss the sending rate tends to grow and eventually overshoots the actual available bandwidth. This causes batch losses and makes the traffic jittery and bursty. Using RRR solves most of these problems. It is a sensitive and smooth estimate of the congestion. Relying on RRR estimates one can take corrective action before packet loss happens. Next we show how such a mechanism can be designed and implemented.

## III. RELATIVE RATE REDUCTION BASED CONGESTION CONTROL

Fair and friendly congestion control should be based on quantities which are the same for each network flow passing through a link. Packet loss probability and round trip time are such quantities, therefore most of the mechanisms rely on them. Since the RRR of a link is determined by the total rate of incoming flows only and its value is the same for each flow we can also build a control mechanism around it. Congestion control mechanisms are designed to keep the network in a steady state. In absence of congestion, packet sending rates of flows are increased, while they are decreased when congestion is observed. This can be done very simply and elegantly in an RRR based framework. Suppose we would like to increase the input rates until the level of congestion – measured by $\hat{p}$ – is below some target value $p_T$. We can define a feedback

algorithm which can adjust the input rate as a function of the output rate by

$$X^{in} = \frac{1}{1 - p_T} X^{out}. \tag{10}$$

Here $X^{in}$ is the actual sending rate of the flow and $X^{out}$ is the measured rate of the received flow. $X^{out}$ is sent back to the the sender with some back-propagation delay. In case the actual level of congestion is $\hat{p}$ the back-propagated value of the observed rate of the received flow is $1 - \hat{p}$ times smaller than the value of the input flow a round trip time ago. So, during one round trip time cycle the input level of the flow is updated approximately by the following macroscopic equation

$$X^{in}(t) = \frac{1 - \hat{p}}{1 - p_T} X^{in}(t - T_{RTT}), \tag{11}$$

where $T_{RTT}$ is the actual value of the round trip time. The input rate is increased or decreased depending on whether the actual congestion is below or above the target level. For a fixed RRR the solution of (11) is exponential $X^{in}(t) = X^{in}(0)e^{\alpha t}$, where the exponent is $\alpha = \log((1 - \hat{p})/(1 - p_T))/T_{RTT} \approx (p_T - \hat{p})/T_{RTT}$. Increasing input rate increases the level of congestion in the system until $\hat{p}$ reaches the target level. The growth of the input rate is stopped and a steady state is reached.

The value of $p_T$ determines the level of congestion in the system. It should be kept as low as possible if the bandwidth and all the other requirements of the flow are met. If the value of $p_T$ is set below the actual congestion level of the system, the flow shrinks exponentially, therefore it is desirable to adjust it properly. Setting $p_T$ to a constant low value produces a flow which is only active when the congestion in the system is very low. Such flows can utilize unconsumed bandwidth in the system. In some applications it can be desirable to set $p_T = 0$ when the bandwidth of the flow is already sufficiently high for the given application. In general all these choices can be realized by making $p_T$ dependent on the rate of the flow at the receiver side

$$p_T = p_T(X^{out}). \tag{12}$$

With this notation the general rate control algorithm reads [1]

$$X^{in} = \frac{1}{1 - p_T(X^{out})} X^{out}. \tag{13}$$

In the next subsection we set up a packet level simulation framework to realize the new rate control algorithm.

### A. Packet Level Realization

We implemented a new rate control algorithm which controls the rate of a standard UDP flow. In this paper we refer this UDP protocol with the introduced rate control algorithm as $RRRP$. The main functionality of RRR is to adjust the packet sending rate via the value of the measured output rate and a specific feedback function. The output rates are measured at the receiver side and sent back to the sender in the acknowledgment packets to utilize it in adjusting the sending rate. The packet sending process is independent from the acknowledgment process. The data packets are injected with the actual sending rate even if no acknowledgment packets arrived. The sending rate is updated when an acknowledgment packet arrives with the back-propagated output rate information. To have a frequent update the receiver sends acknowledgment for each data packets. Between updates the sending rate is kept constant. The update is based on the specific $p_T(X^{out})$ function applied in the sender. The measured output rate is averaged for $n$ consecutive packets to avoid undesirable synchronization. Besides the averaged quantities in the following we use the same symbols for input and output rate as before

$$X^{out} := \langle X^{out} \rangle_n. \tag{14}$$

[1] In case $X^{in}$ exceeds the maximal capacity of the sender $C^{max}$, then $X^{in}$ should be set to $C^{max}$ instead of the value (13).

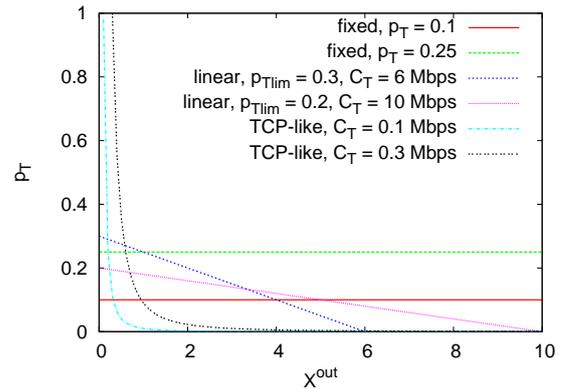

Fig. 1. The investigated feedback functions: feedback with fixed $p_T$, with linear relation between $X^{out}$ and $p_T$, and a "TCP-like" function where $p_T \sim (X^{out})^{-2}$. All kinds of functions are plotted with two sets of parameters.

The effects on congestion control of using different $p_T(X^{out})$ feedback functions were investigated in packet level simulations. The simulations were performed in NS-2 [24] in a dumbbell topology shown in Figure 2.

In the next sections we investigate some $p_T(X^{out})$ feedback functions and we discuss their most important consequences. We focus on the following performance characteristics: optimal utilization of the link capacity, intra-protocol fairness and TCP friendliness.

## IV. RATE DEPENDENT FEEDBACK FUNCTIONS

The investigated three important feedback functions. The constant function for pedagogical reason. A linear function which is ideal for utilization unconsumed network resources. A "TCP like" function, which can adapt network conditions as effectively as TCP does. These functions are shown in Figure 1 with different parameters.

The performance of three different feedback functions are tested in a topology shown in Figure 2. The physical bandwidths for the access links are $C = 100 Mbps$, for the bottleneck link is $C = 10 Mbps$, the buffer length on the bottleneck link is $B = 50 pkt$, while the link delays are set corresponding to the appropriate simulated scenario. In this section we used the $d = 10 msec$ link delay on each links. For background traffic we used fixed size packets ($P = 12000bit$) with Poisson arrival process. In this section we present scenarios where the background traffic is fluctuating due to its arrival process, and also sudden changes occur during the simulation at $t = 50, 90$ and $120sec$.

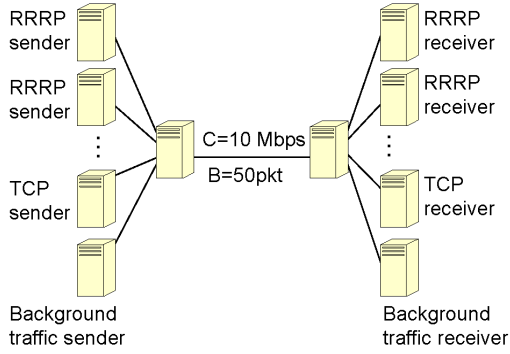

Fig. 2. Topology used in all the simulated scenarios. The physical bandwidth for the access links is $C = 100Mbps$, for the bottleneck link is $C = 10Mbps$, while the link delays are set according to the appropriate simulated scenario.

*A. Fixed $p_T(X^{out})$*

One of the most basic feedback functions is the constant function,

$$p_T(X^{out}) = p_{fix}, \quad (15)$$

where $p_{fix}$ is our preset constant parameter. This parameter determines the target congestion which the protocol would like to reach. If the actual $\hat{p}$ is smaller than the preset constant, then the control algorithm increases the sending rate, while if it is higher, then the sending rate will decrease exponentially. This feedback mechanism results a quick and tight adaptation to the variable available bandwidth conditions. These can be seen in Figure 3a and b, where the throughput and the observed packet loss rate are plotted. In the simulation the target RRR is set to $p_{fix} = 1\%$ and the initial bandwidth is set to $X^{in} = 1.6Mbps$. The RRR controlled flow starts at $t = 20sec$ and adapts to the available bandwidth. The input rate increases until the RRR of the flow reaches $\hat{p} = 1\%$ the value of the preset $p_{fix}$. The $\hat{p}$ falls down to zero at $t = 50sec$, since bandwidth suddenly becomes available and the flow reaches the receiver without rate change. From this time the input rate increases until $\hat{p}$ reaches $p_{fix}$ again. The same process happens again at $t = 90sec$, when the background traffic switched off. At $t = 120sec$ a high bandwidth background traffic is switched on and the value of $\hat{p}$ jumps high. The input rate of the flow is set to the decreased available bandwidth instantaneously by the control algorithm. In Figure 3b the observed packet loss probabilities can be seen. It follows the preset $p_{fix}$ during the stationary intervals, while it falls to zero if more available bandwidth appears and jumps high when the background traffic increases.

This feedback mechanism keeps the $p_{fix}$ target RRR even when it is unnecessary, for example when the flow utilizes the total bandwidth of the system already. To avoid this, next we introduce a feedback function, where $p_T$ decreases to zero proportionally with the distance of the actual $X^{out}$ and the preset $X^{out}_{Tlim}$ limit.

*B. Linear $p_T(X^{out})$*

A simplest relation between $p_T$ and $X^{out}$ can be written

$$p_T(X^{out}) = p_{Tlim} \cdot \left(1 - \frac{X^{out}}{X_{Tlim}}\right), \quad (16)$$

where $p_{Tlim}$ is the base value of the target RRR and $X_{Tlim}$ is the maximum rate where we would like to limit the speed of the transfer. The $p_{Tlim} = 0$ for output rates exceeding the limit $X_{Tlim}$. The parameter $p_{Tlim}$ defines the maximum level of congestion where the protocol is still able to work, for higher level of congestion it stops. The limit $X_{Tlim}$ can be regarded as the maximum bandwidth requirement of an application.

The main properties of RRRP with linear feedback function is shown in Figure 3c and d. In the simulation the packet loss limit is set to $p_{Tlim} = 1\%$, the maximal bandwidth limit $X_{Tlim} = 4Mbps$, and the initial bandwidth $X^{in} = 1.6Mbps$. As the measured output rate is getting closer and closer to its limit the target congestion level and the observed packet loss rate is decreasing due to the linear relation between $p_T$ and $X^{out}$. As the output rate reaches its $X_{Tlim}$ limit at $t = 50sec$ the output bandwidth is kept $X^{out} = X_{Tlim}$ even when there is more available bandwidth in the network path. In the interval between $50sec$ and $120sec$ the output rate remains $X^{out} = X_{Tlim}$ and the observed packet loss rate is zero.

The advantage of this feedback function is that we could develop a control algorithm which utilizes the network only if the level of congestion is below a preset $p_{Tlim}$ level and the consumed bandwidth never exceeds $X_{Tlim}$. This can be ideal for an application which would like to utilize only unconsumed resources of the network.

The drawback of this mechanism is that it does not provide a generic adaptation scheme, which could operate at a wide range of network conditions like TCP.

*C. "TCP like" Feedback Function*

To develop a sufficiently generic control mechanism, which can adapt to the network conditions the way TCP does in current networks we rely on the macroscopic properties of TCP flows.

The average bandwidth utilized by a TCP flow

$$X^{out}_{TCP} = \frac{k \cdot P}{T_{RTT}\sqrt{p_{TCP}}}, \quad (17)$$

where $P$ is the data packet size, $p_{TCP}$ is the packet loss probability, $T_{RTT}$ is the round trip time, while $k$ is a constant value between 1 and 2 [23]. We can invert this relation and can express the loss rate as a function of the current bandwidth utilization

$$p_{TCP} = \left(\frac{k \cdot P}{T_{RTT} \cdot X^{out}_{TCP}}\right)^2. \quad (18)$$

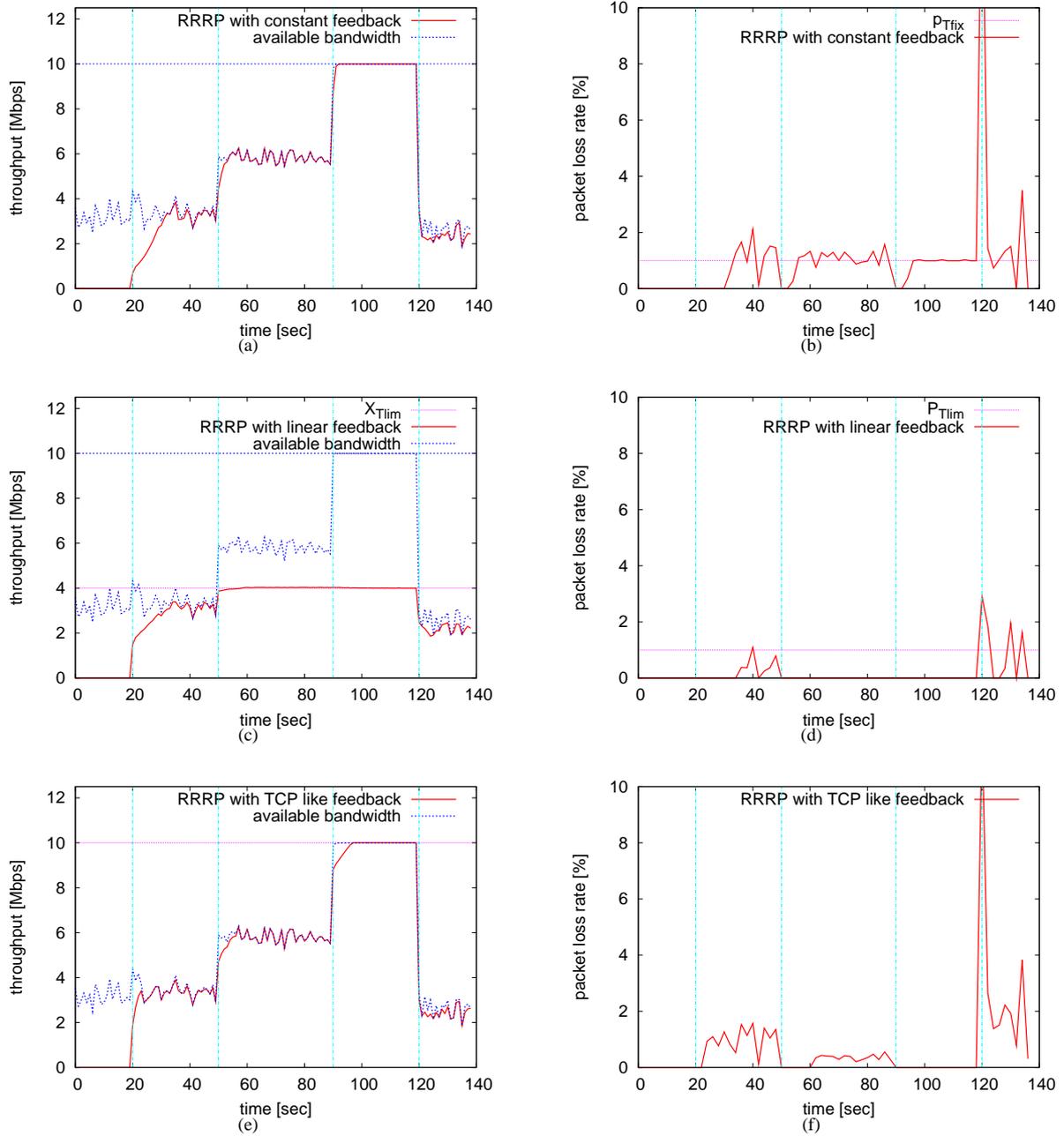

Fig. 3. Adaptation of RRRPs to variable background traffic. The three rows represent the results for RRRPs with the three discussed feedback functions (top - fixed, middle - linear and bottom - "TCP like"). The used parameters are: for the fixed feedback $p_T = 1\%$, for the linear feedback $p_{Tlim} = 1\%$, $X_{Tlim}^{out} = 4 Mbps$, and for the "TCP like" $X_T = 0.3 Mbps$. In the right column the throughput of the RRRP flow and the available bandwidth left by the backgound traffic can be seen, while the average packet loss rates can be seen in the left column.

We can regard this expression as a fixed connection of $X^{out}$ utilized bandwidth and the $p$ packet loss probability. One can define a feedback function with $p_T(X^{out})$ such that it mimics the packet loss rate of the corresponding TCP with the same received bandwidth $X^{out}$. This would result a TCP friendly control mechanism, since the sending rate is the same as it would for a TCP flow at the same packet loss conditions. In general we can define the feedback function as

$$p_T(X^{out}) = min\left(\left(\frac{X_T}{X^{out}}\right)^2, 1\right), \quad (19)$$

where $X_T$ is a tunable parameter.

The Eq.(18) connects the $X^{out}$ utilized bandwidth to the $p_{TCP}$ packet loss rate in a fixed manner, while (19) gives the possibility to adjust the relation via the parameter $X_T$. The advantage of adjusting this parameter is discussed in detail in the next section.

The shape of the (19) allows that flows can work at any packet loss rate, while their expected congestion level is decreasing with the increasing bandwidth utilization. This version keeps sending data packets at any network conditions with an appropriate rate. The mechanism guarantees the quick adaptation to the available bandwidth in the network path, while it can also work at slow speeds. These properties are presented in Figure 3e and f, where the the $X_T$ parameter is set to $0.3Mbps$, and the initial bandwidth is set to $X^{in} = 1.6Mbps$. The flow quickly adapts to the available bandwidth and tightly follows it, due to the applied feedback function. The packet loss rate depends on the received bandwidth, which can be seen in the Figure 3f. At $t = 120sec$ the high peak in the packet loss is the aftermath of the sudden background traffic increase in the network as usual.

As we show next this feedback function also leads to stable bandwidth sharing between flows.

## V. Performance Evaluation

A congestion control has to satisfy several different requirements to achieve good overall network performance. The general rate control algorithm defined in (13) inherently provides exponential adaptation to available bandwidth achieving high throughput. This feature have been already discussed in previous section. Next we show that the control algorithm itself guarantees the fairness between the same kinds of protocols, it is friendly behaviour against other kind of flows and the throughput of the flows are sufficiently smooth to let streaming application work with small buffering. Other important properties like the CPU usage, required computation power and initial parameter settings are not investigated here.

### A. Intra-Protocol Fairness

Designing an adaptive flow control an important goal is to fairly distribute the bandwidth between the competing flows. The intra-protocol fairness is measured between the same kind of protocols with the same kind of congestion control algorithm. Usually a single bottleneck case is considered, where the bandwidth should be shared in equal portions between the flows in fair situation.

In this subsection we investigate the intra-protocol fairness of our RRR based control algorithms. Flows with fixed $p_T$ feedback mechanism cannot maintain a stationary bandwidth sharing, since their prescribed congestion level is not adaptive in terms of the output rate. The linear feedback mechanism can be adaptive only in a restricted bandwidth and congestion level range. Only the "TCP like" feedback mechanism can be considered as a congestion control for a general use, and will be discussed here.

In Figure 4 the results for four competing flows, sharing a single bottleneck link, can be seen. The physical bandwidth of the bottleneck link is $C = 10Mbps$. In the upper row of Figure 4 a simulated scenario can be seen, where the links have the *same* propagation delay $d = 1msec$ in the topology of Figure 2. The lower row shows the results of an other scenario, where the link delays are significantly *different* from each other. In this case the link delays were $d_i = 1, 10, 50$ and $100msec$ for the different flows. The topology was the same. In both scenarios we used "TCP like" feedback functions with $X_T = 0.6Mbps$ and the initial bandwidth was set to $X^{in} = 1.6Mbps$. The senders start their session at $0, 20, 40$ and $60sec$ respectively, while they stop sending at $100, 120, 140$ and $160sec$. In the left side the received bandwidth share is shown for all the flows. For both scenarios the bandwidth is fairly shared among the flows even in scenario with different link delays. This is because the feedback function is not sensitive to the round trip time, unlike the TCP congestion control algorithm, which is drastically unfair for flows with different RTT values. The convergence to the fair share is very quick for all the cases. The difference between the two scenarios can be seen only in the little bit slower adaptation of the flows with higher link delays. This is because their feedback have much longer back-propagation delay than the flows with low link delays. The sum of the throughputs of individual flows is also presented with dashed line in Figure 4a and d, which shows that the link utilization is very high.

In the right side of Figure 4 the packet loss rates can be seen. The packet loss rates increase as more and more flows share the same link, since the packet loss rates are higher (see Eq.19) for smaller $X^{out}$ values, corresponding to the fair bandwidth share. Here we would like to note that this phenomenon can be observed also in case of TCP flows sharing a common link.

We can conclude that the RRR based congestion control with the same feedback functions and parameters results fair resource sharing in network paths. This optimal sharing is reached very quickly, in a few round-trip-times. Each flow receives the same amount of bandwidth, suffer the same queuing delay at the bottleneck buffer, and also their packet loss rates are equal. These properties remain valid even if the propagation delay is not equal for the different flows sharing the same bottleneck link.

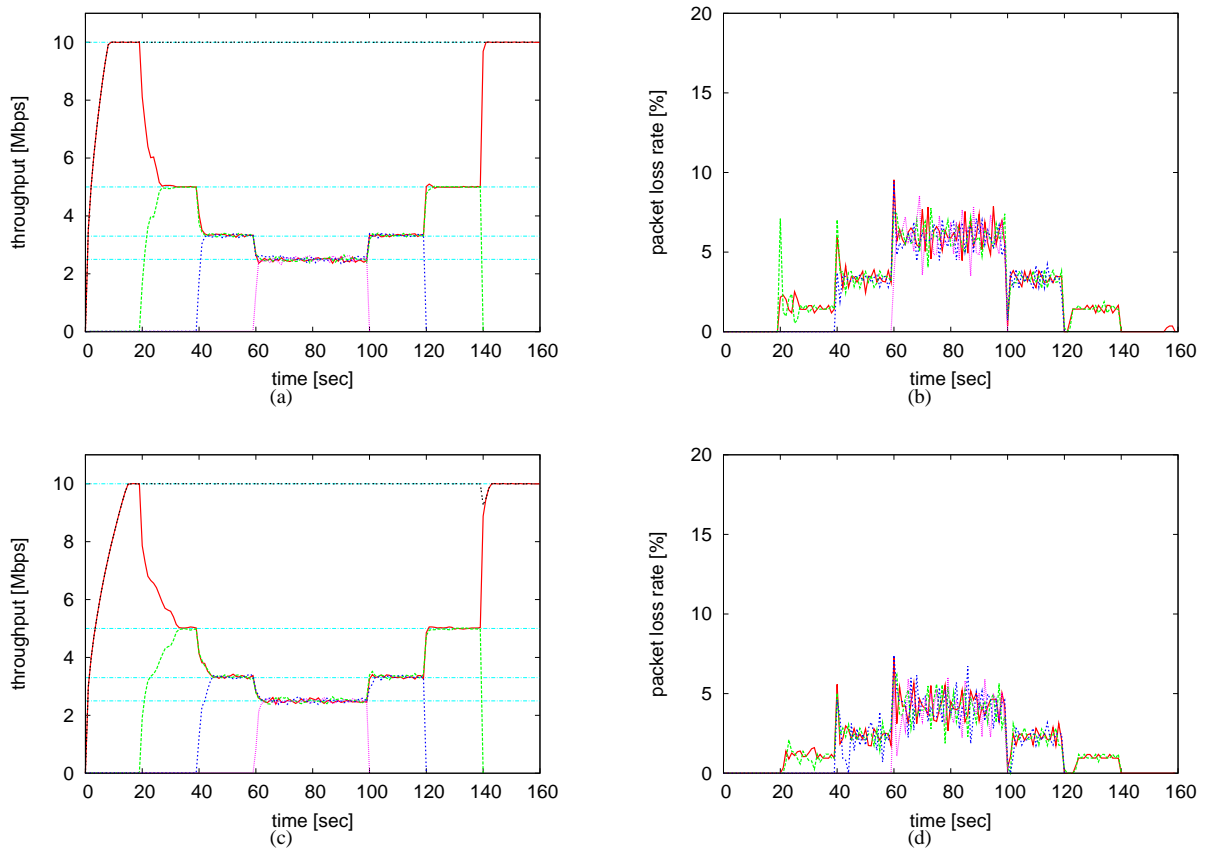

Fig. 4. Main properties of four competing RRR flows with "TCP like" feedback function. In the **upper row** a scenario with the same link delays ($1msec$ for each) can be seen, while in the **lower row** the access link delays for the competing flows are significantly different ($1, 10, 50, 100msec$), while the delay of the common link is $1msec$. In the **left side** the throughput can be seen, the fair shares are also indicated with dotted lines. In the **right side** the packet loss probabilities can be seen.

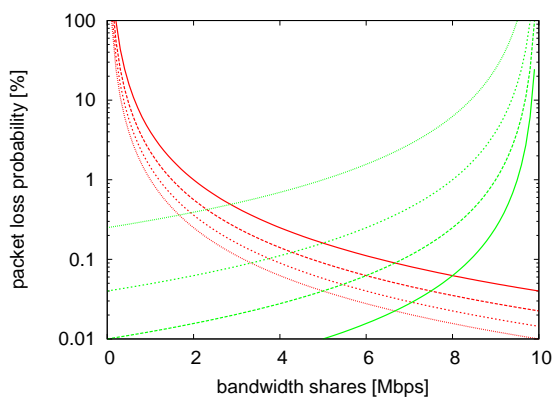

Fig. 5. Loss-bandwidth function of TCP and the "TCP like" feedback functions with several $X_T$ parameters. The intersection represents the stable bandwidth sharing between regular TCPs and the proposed flow. The curves are based on the macroscopic fluid approximation.

### B. TCP Friendliness

In this section we would like to present how a conventional TCP version and the RRR based control method with a "TCP like" feedback function can share the network resources and how the $X_T$ parameter determines the bandwidth sharing of TCP and RRRP flows. First, we would like to determine the stable operating point of TCP and RRRP flows on a bottleneck link with physical capacity $C$ ($C = 10Mbps$ in our examples). For sufficiently long time the packet loss rate in the operating point and $\hat{p}$ should coincide, therefore in the operating point $p_T(X_{RRRP}^{out}) = p_{TCP} = p^*$. Denoting the stable bandwidth share of the TCP flow by the $X^* = X_{TCP}^{out}$, the bandwidth share of RRRP flow is $X_{RRRP}^{out} = C - X^*$.

In Figure 5 we show a graphical solution for the operating point. The curves decreasing from left to right are representing the TCP curves (Eq.(18)) and the curves decreasing from right to left are the "TCP like" feedback functions (Eq.(19)) plotted in reverse direction from the physical bandwidth $C = 10Mbps$. In this way the intersection $(X^*, p^*)$ of two specific curves represents the stable bandwidth share of TCP and the corresponding packet loss probability.

Next, we study three scenarios how the stable bandwidth share sets up between real RRRP and TCP flows. The topology of the simulated scenarios is shown in Figure 2, with bottleneck physical bandwidth $C = 10Mbps$, and with the same link delays for all the links $d = 10ms$. The maximum buffer length is $B = 50 pkts$, and the initial bandwidth is set to $X^{in} = 1.6 Mbps$. In Figure 6 results are shown for three scenarios with different $X_T$ parameters. The parameters of the TCP, such as packet loss and RTT, are determined by the network conditions without our intervention. In all scenarios the TCP flow starts at $t = 20 sec$ and the RRRP flow starts at $t = 50 sec$. Due to page limitation, we plot the throughput of both flows and the total bandwidth utilization together with the observed packet loss rates in the left column of Figure 6.

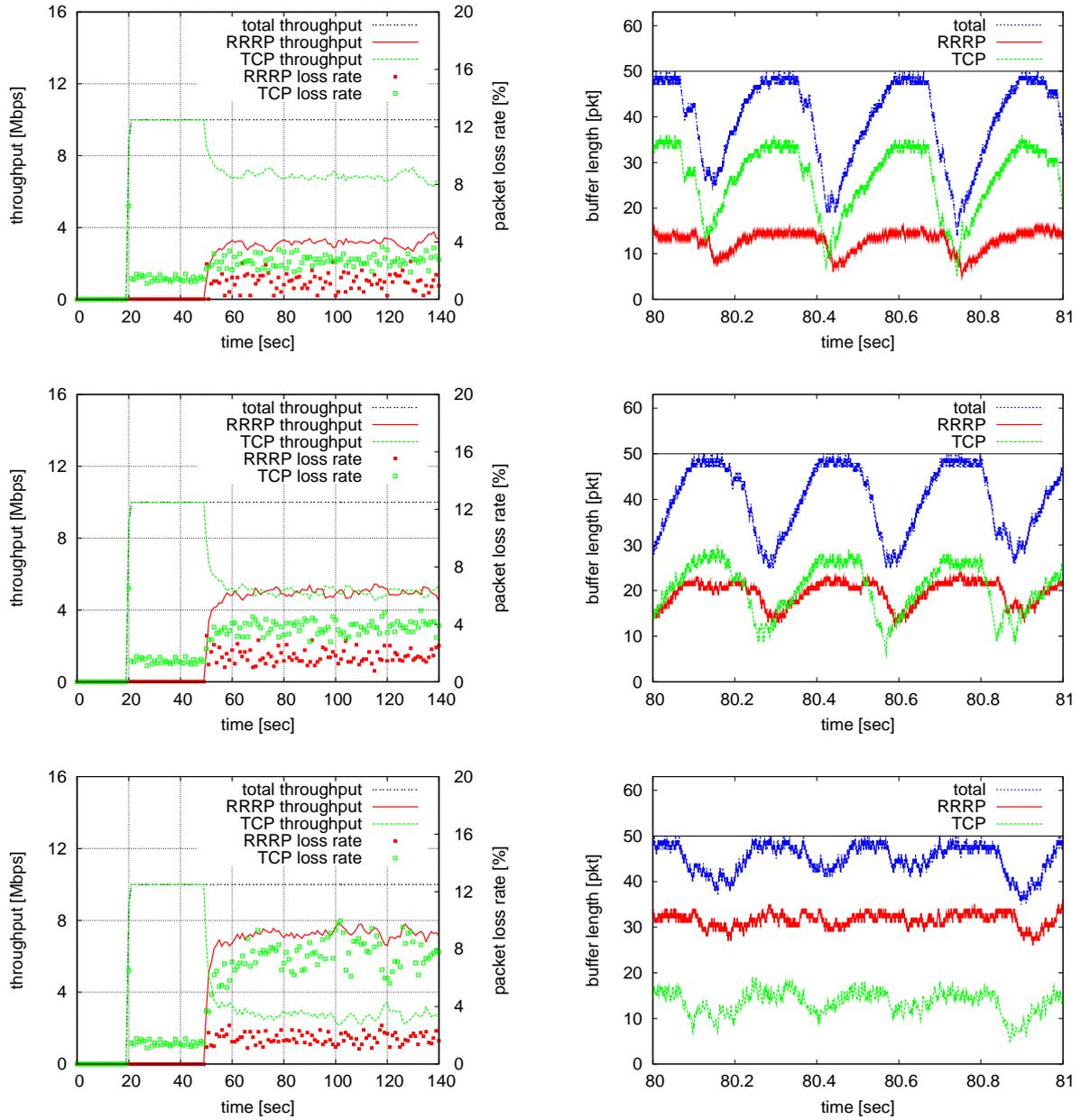

Fig. 6. **Left column**: Bandwidth sharing for conventional TCP-Reno and a RRR controlled UDP flow (with solid lines, left axis) and the packet loss probabilities (with symbols, right axis). The RRR controlled flow has "TCP like" feedback function with different $X_T$ parameters. The parameters used are the following: in the **top row** $X_T = 0.3 Mbps$, in the **middle row** $X_T = 0.6 Mbps$ and in the **bottom row** $X_T = 0.9 Mbps$. In the **right column** the typical buffer lengths are shown.

The numerical values of throughput can be read on the left axis in $Mbps$, while the packet loss percentage can be seen in the right axis. The total buffer length and the number of buffered packets for both flows are shown in the right column. The top row shows results for $X_T = 0.3 Mbps$ (fixed during the simulation). A stable bandwidth sharing between the flows can be seen. The preset $X_T$ value determines the portions of received bandwidths of the flows. In this case the TCP flow receives more bandwidth than the RRRP flow. The packet loss rate is a little bit higher for the TCP flow, since its packet sending is more bursty which leads to batch packet losses is the case when the length of the queue is close to the maximum buffer size. The queue length shows sawtooth pattern, since the TCP utilizes most of the bandwidth. In the middle row the $X_T$ is set to $0.6Mbps$. The flows share the bandwidth equally in this case. The buffer utilization is equal for the flows, but it is clear that the number of queued packets for the RRRP is much smoother, than that of TCP. In the bottom row the $X_T$ is set to $0.9Mbps$. In this case the RRRP flow receives more bandwidth. The packet loss rate is significantly higher for the TCP flow, as packets sent in bursts by the TCP often suffer batch packet losses due to the almost saturated buffer.

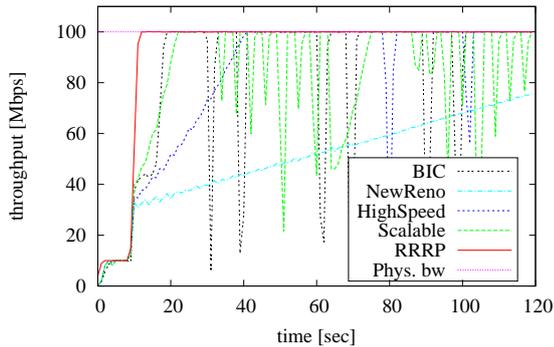

Fig. 7. Throughput of different high-speed TCP variants and RRRP in an up-switch scenario. The up-switch in the physical bandwidth from $C = 10Mbps$ to $C = 100Mbps$ happens at $t = 10sec$.

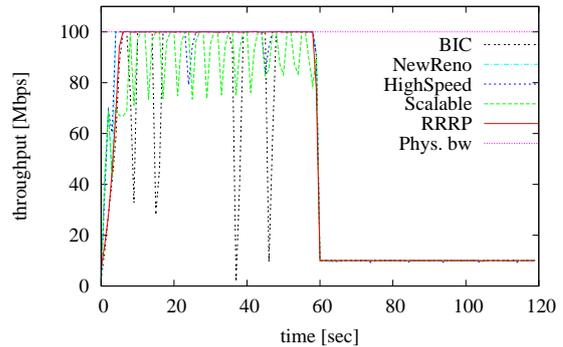

Fig. 8. Throughput of different high-speed TCP variants and RRRP in a down-switch scenario. The down-switch in the physical bandwidth from $C = 100Mbps$ to $C = 10Mbps$ happens at $t = 60sec$.

We can conclude that with different parameter settings our congestion control algorithm with "TCP like" feedback function has adjustable friendliness to conventional TCP flows. It is also clearly seen that for all the cases the number of buffered packets for the RRRP is much smoother, than that is for TCP. This is due to the smoothness of the rate based sending process. The total bandwidth utilization is very close to the physical capacity of the links for all the three scenarios, which demonstrates that RRRP exploits efficiently the residual capacity left by the sawtooth pattern of TCP.

## VI. TESTBED EXPERIMENTS

We implemented a UDP-based prototype of our congestion control algorithm in C++ on standard Linux operating system to validate its performance in real heterogeneous hardware environment. Our prototype is not a complete protocol but it is suitable to investigate the control algorithms and perform real world tests including real reliable file transfer. A future goal is to implement the missing parts of a full protocol stack: flow control, socket API, etc. We also plan to create a Linux kernel module version of the RRRP. Since our rate control does not rely on packet loss events it is robust against even very high packet loss rates ($\gg 1\%$). To cope with the high packet loss rates we implemented a new retransmission algorithm. It runs completely separately from the rate control and they can be developed independently. To support high loss resistance we use acknowledgments with much more loss information than cumulative acknowledgments have. Typically our acknowledgment contains a full-report on which packet has arrived at the receiver side and which has not. This is much in flavor of the TCP SACK option [25] and the UDT NAK [21]. In principle we could handle very high loss rates by allowing full-report acknowledgments. In practice we have to limit the network overhead of the acknowledgment flow so we use acknowledgment packets with less information.

Our testbed consists of standard PCs with gigabit NICs and switches. All the network properties are emulated with a kernel-module. The emulator is capable to include drop-tail queuing, service rate adjustment, propagation delay and packet loss settings as well as mobile handover emulation. The experiments performed in wired scenarios validated the performance already shown in simulated scenarios of Section IV and V. Here we present only our testbed experiments relevant in mobile environments.

One of the most challenging environments for a transport protocol is a wireless network. In [26] the authors investigate the impact of the evolving wireless technologies on the performance of different high-speed transport protocols (HighSpeed TCP [6], Scalable TCP [5], BIC TCP [27]) and the NewReno TCP [3]. They set up some emulated scenario which represents some of the challenges in wireless and mobile environments. In wireless networks the inter-system handovers are characteristic events. We distinguish the up- and down-switch handovers. Up-switch means to change from a "low-speed" to a "high-speed" technology (e.g. from HSPA to LTE) during the transmission, while the down-switch is vica versa. In the testbed the "low-speed" technology was emulated with a $10Mbps$ link and the "high-speed" one with a $100Mbps$ link, in accordance with the real parameters. We repeated the experiments of [26] with our "TCP like" congestion control algorithm to compare its performance to different high-speed transport protocols. Each protocol was tested separately and they are plotted in the same figure for demonstration only.

In Figure 7 the throughputs of the protocols are shown for an up-switch scenario. The switch happened at $t = 10sec$, when the physical capacity of the route suddenly changed from $C_1 = 10Mbps$ to $C_2 = 100Mbps$. Protocols start to grow up right after the change. As we can see RRRP has the fastest (exponential) adaptation rate, while TCP versions are hampered by their slow additive increase mechanisms. In Figure 8 a down-switch scenario can be seen. The switch happened at $t = 60sec$, when the physical capacity dropped. In this case all the protocols adapt immediately to the lower bandwidth successfully. The bandwidth usage of the RRRP is stable, while the throughputs of the TCP versions fluctuate.

## VII. Conclusion

In this paper we introduced a new way of congestion control based on the relative rate reduction replacing the erratically changing packet loss. We studied three different feedback functions. One of them was suitable to harness available bandwidth without congesting the network path. We demonstrated that the other "TCP like" version is fair and TCP friendly, while it has superior adaptation properties. It utilizes optimally the bandwidth of a network path in presence of both elastic and inelastic background traffic. It adapts to the sudden bandwidth changes much faster than high-speed TCP variants. We validated the main concepts of the RRRP in real testbed experiments. We hope that with a robust acknowledgment scheme and with time dependent feedback $p_T(X^{out}, t)$ RRRP could become a mainstream protocol for Future Internet applications.

## Acknowledgment

We thank for the fruitful discussions to András Veres and Gábor Németh and for the NS code to Attila Fekete. The authors thank the partial support of the National Office for Research and Technology (NKFP 02/032/2004 and NAP 2005/KCKHA005).